\documentclass[manuscript]{aastex}

\shorttitle{Quiescent galaxies at z$\sim$3}
\shortauthors{Fan et al.}

\begin{document}

\newcommand{\kms}{\>{\rm km}\,{\rm s}^{-1}}
\newcommand{\reff}{r_{\rm{eff}}}
\newcommand{\msol}{M_{\odot}}
\newcommand{\gf}{{\tt GALFIT}~}
\newcommand{\mh}{H_{\rm{F160W}}}
\newcommand{\mj}{J_{\rm{F125W}}}
\newcommand{\mi}{I_{\rm{F814W}}}
\newcommand{\mv}{V_{\rm{F606W}}}
\newcommand{\ser}{S\'ersic~}
\newcommand{\sext}{{\tt SExtractor}~}

\title{THE STRUCTURE OF MASSIVE QUIESCENT GALAXIES AT Z$\sim$3 IN THE CANDELS-COSMOS FIELD}

\author{Lulu Fan \altaffilmark{1,2}, Guanwen Fang\altaffilmark{3,4},Yang Chen\altaffilmark{5,1}, Zhizheng Pan\altaffilmark{1,2},Xuanyi Lv\altaffilmark{1,2}, Jinrong Li\altaffilmark{1,2}, Lin Lin\altaffilmark{1,2}, Xu Kong\altaffilmark{1,2}}
\altaffiltext{1}{Center for Astrophysics, University of Science and Technology of China, 230026 Hefei, China}
\altaffiltext{2}{Key Laboratory for Research in Galaxies and Cosmology, USTC, CAS, 230026, Hefei, China}
\altaffiltext{3}{Institute for Astronomy and History of Science and Technology, Dali University, Yunnan,671003, China}
\altaffiltext{4}{Key Laboratory of Modern Astronomy and Astrophysics, Nanjing University, Ministry of Education, Nanjing 210093, China}
\altaffiltext{5}{Astrophysics Sector, SISSA, Via Bonomea 265, 34136 Trieste, Italy}

\begin{abstract}
In this letter, we use a two-color ($J-L$) vs. ($V-J$) selection criteria to search massive, quiescent galaxy candidates at $2.5\leq z \leq 4.0$ in the CANDELS-COSMOS field. We construct a $\mh$-selected catalogue and complement it with public auxiliary  data. We finally obtain 19 passive VJL-selected (hereafter pVJL) galaxies as the possible massive quiescent galaxy candidates at $z\sim 3$ by several constrains. We find the sizes of our pVJL galaxies are on average 3-4 times smaller than those of local ETGs with analogous stellar mass. The compact size of these $z\sim3$ galaxies can be modelled by assuming their formation at $z_{form}\sim 4-6$ according to the dissipative collapse of baryons. Up to $z<4$, the mass-normalized size evolution can be described by $r_e\propto  (1+z)^{-1.0}$. Low \ser index and axis ratio, with median values n$\sim1.5$ and $b/a\sim0.65$ respectively, indicate most of pVJL galaxies are disk-dominated.
Despite large uncertainty, the inner region of the median mass  profile of our pVJL galaxies is similar to those of quiescent galaxies (QGs) at $0.5<z<2.5$ and local Early-type galaxies (ETGs). It indicates local massive ETGs have been formed according to an inside-out scenario: the compact galaxies at high redshift make up the cores of local massive ETGs and then build up the outskirts according to dissipationless minor mergers.
\end{abstract}


\section{Introduction}

Early-type galaxies (ETGs) are the most massive objects in
the local Universe, containing the bulk of the stellar mass, which are predominantly old with mass-weighted ages of $\geq 8-9$ Gyr (Renzini 2006). This indicates that most of the stars in ETGs were formed at redshift $z \geq 1.5$. ETGs are characterized by homogeneous stellar populations with a lot of observational scaling relations. Amongst these, a remarkably tight luminosity-size/mass-size correlation has been confirmed (e.g. Shen et al. 2003).

Recent observations have evidenced that massive quiescent galaxies at high redshift are more compact compared to local early-type galaxies with similar stellar mass (Daddi et al. 2005; Trujillo et al. 2006; Toft et al. 2007; van der Wel et al.2008; van Dokkum et al. 2008; Damjanov et al. 2009; Ryan et al. 2012; Papovich et al. 2012; Zirm et al. 2012). Physical mechanisms have been proposed to explain the smaller size at high redshift and the resulting size evolution with redshift, such as major merger, dissipationless  (dry) minor merger  (e.g. Naab et al. 2009), ``puff-up" due to the gas mass loss by AGN  (Fan et al. 2008) or supernova feedback  (Damjanov et al. 2009). 

So far, most studies on the structural evolution of passive galaxies focus on the redshift range $0<z<3$. Although the massive quiescent galaxies are rarely selected at $z\geq 3$, the discovery of these galaxies are quite exciting.
Several  methods (such as morphological, color-color or specific star formation rate (sSFR); Cassata et al. 2013, Bruce et al. 2012, Szomoru et al. 2012) for selecting high redshift quiescent galaxies have been used.  For instance, the rest-frame UVJ color has been shown to effectively separate quiescent galaxies from star-forming galaxies (e.g. Williams et al. 2009; Patel et al. 2012). Anyway, the rest-frame UVJ color and sSFR are dependent on the determination of photometric redshift. Although photometric redshift now can be measured with relatively small error (e.g. Ilbert et al. 2009) , the measurement of photometric redshift may vary from person to person, depending on the adopted SED-fitting codes and SED libraries. Guo et al. (2012) proposed a new set of color selection criteria analogous with the BzK method (Daddi et al. 2004) to select both star forming galaxies and quiescent galaxies at $z\sim3$. They extended the successful BzK method from $z\sim2$ to $z\sim3$ by replacing the selection bands with the \emph{observed} V, J, and IRAC 3.6 $\mu m$ band (hereafter L-band), according to the relative shift of galaxy spectra between the two redshifts. Unlike photometric redshift selections, the bias of color selection can be fairly explicitly determined and the results are robust.

In this letter,we will search for massive quiescent at $z\sim3$ using the new VJL color selection method. We will use recent HST/WFC3 imaging on the central region of the Cosmic Evolution Survey (COSMOS) as part of CANDELS multi-cycle treasury programme (Grogin et al. 2011; Koekemoer et al. 2011) to study the structure of the two-color selected massive, quiescent galaxies at $z\sim3$. We will try to extend our knowledge on size distribution and  evolution with redshift  beyond $z\sim 3$. Throughout this letter, we assume a concordance $\Lambda$CDM cosmology with $\Omega_{\rm m}=0.3$, $\Omega_{\rm \Lambda}=0.7$, $H_{\rm 0}=70$ $\kms$ Mpc$^{-1}$. All magnitudes are in the AB systems.

\section{Data and VJL color selection}

We focus our study on the central region of the COSMOS survey (Scoville et al. 2007) which has been imaged with HST/WFC3 as part of CANDELS multi-cycle treasury programme (Grogin et al. 2011; Koekemoer et al. 2011). The CANDELS data consist of a contiguous mosaic of $4\times 11$ HST WFC3/IR tiles covering a total area of $\sim 210$ arcmin$^2$, along with a contemporaneous mosaic of ACS parallel exposures. The exposure time are 1650, 3450, 1000 and 1600 seconds for F606W ($\mv$ band), F814W ($\mi$ band), F125W ($\mj$ band) and F160W ($\mh$ band) filters, reaching $5-\sigma$ point-source sensitivity of 28.1, 28.0, 26.8 and 26.5 (AB mag) respectively.  The HST WFC3/IR and ACS images are prepared by drizzling the individual exposures onto a grid with rescaled pixel sizes of 60 mas and 30 mas, respectively (Koekemoer et al. 2011). We use the latest data release v1.0 of CANDELS-COSMOS field \footnote{http://candels.ucolick.org/data$\_$access/Latest$\_$Release.html} .

We construct a $\mh$-selected catalogue using \sext (Bertin \& Arnouts 1996) v2.8.6. We run \sext in dual mode using $\mh$ band mosaics as the detection images and ACS $\mv$,$\mi$ and WFC3 $\mj$,$\mh$ bands as the measurement images. We use the same set of \sext detection parameters as the so-called ``hot" setup in van der Wel et al. (2012), which is chosen for optimally detecting small, faint objects. We use a Gaussian smoothing kernel with a FWHM of 4 pixels and select objects with 10 adjacent pixels with 0.7$\sigma$ fluxes.  Deblending is done with a minimum contrast of 0.001 and 64 logarithmic sub-thresholds. Finally, we obtain a $\mh$-selected catalogue of 34,963 objects with $\mh \leq 26.5$.

We complement the $\mh$-selected catalogue with Subaru BVgriz (Capak et al. 2007), UltraVISTA YJH$K_s$ photometry (McCracken et al. 2012) and deep Spitzer/IRAC 3.6 $\mu m$, 4.5 $\mu m$, 5.8 $\mu m$ and 8.0 $\mu m$ data taken from S-COSMOS survey (Sanders et al. 2007). The $\mh$-selected catalogue has been cross-matched with these photometric catalogues, with a matching radius of $1"$. The final catalogue includes multi-wavelength photometry spanning from B band to 8.0 $\mu m$.  The photometric redshifts are determined using \texttt{EAZY} code (Brammer, van Dokkum \& Coppi 2008). The galaxy physical properties, such as stellar masses ($M_*$), star formation rates and luminosity-weighted ages, are derived using \texttt{FAST} (Kriek et al. 2009). We adopt a grid of Bruzual \& Charlot (2003) models assuming a Chabrier (2003) IMF, solar metallicity, exponentially declining star formation histories (SFH) and Calzetti extinction law (Calzetti et al. 2000). We find that the derived $M_*$ don't significantly depend on the assumed SFH.  The difference between the derived $M_*$ using exponentially declining and truncated SFH is only 0.06 dex.

\begin{figure}
\plotone{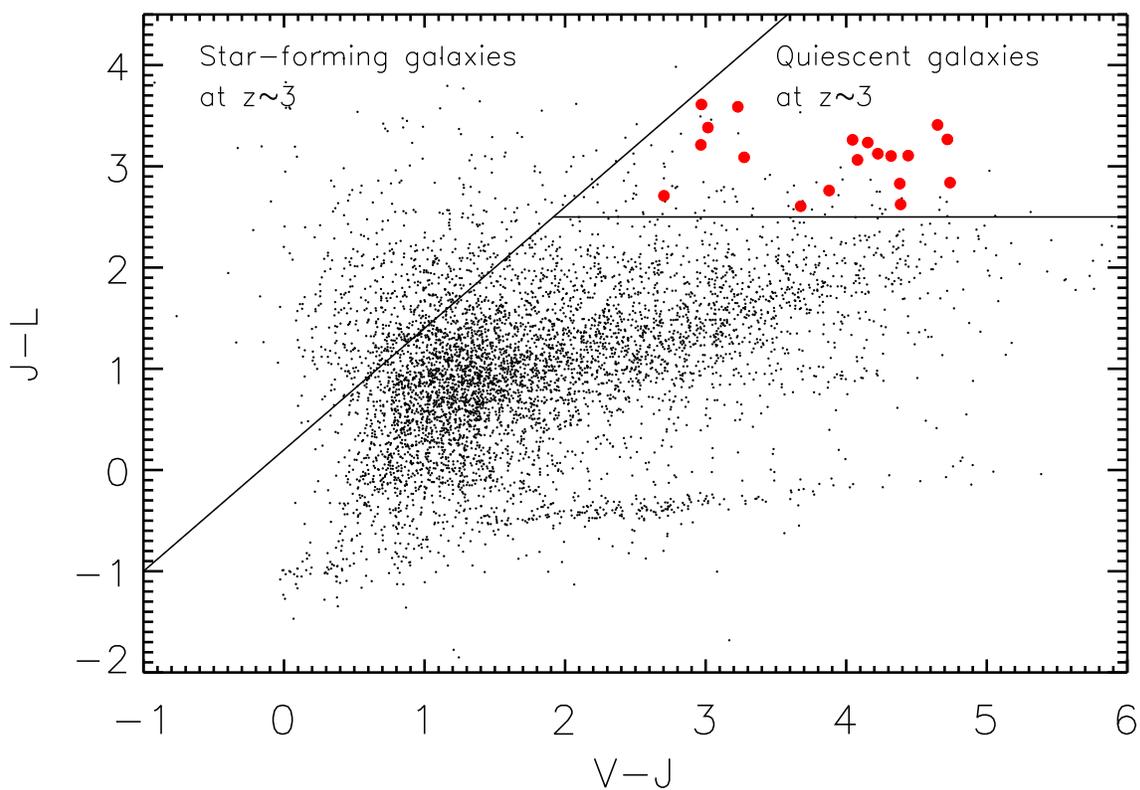}
\caption{Two-color ($J-L$) vs. ($V-J$) diagram for the galaxies in the CANDELS-COSMOS field. Two black lines show the color selection criteria: upper left for sVJL and upper right for pVJL. The small dots show the galaxies in our $\mh$-selected catalogue with $\mh\leq 24.5$. The filled points are the final pVJL candidates.}
\end{figure}

We use pVJL color selection criteria given by Guo et al.(2012):
\begin{equation}
J-L\geq2.5 \bigwedge J-L < 1.2\times(V-J)+0.2
\end{equation}
where $\bigwedge$ means the logical \textit{and}.
In Figure 1, we plot the galaxies of our master catalogue in two-color ($J-L$) vs. ($V-J$) diagram with small black dots. The black lines mark the adopted pVJL color selection criteria (upper right region).  276 galaxies fulfill the color selection criteria. However, just as mentioned in Guo et al. (2012), several possible contaminations can enter the pVJL selection window.  The main contamination may be the highly obscured star forming galaxies (SFGs) from both lower ($z\leq 2.0$) and higher ($z\geq 4.0)$ redshifts. And active galactic nucleus (AGNs) could also contaminate our pVJL sample. We therefore choose several conditions to constrain and clean our pVJL sample for the purpose of studying the structure of massive, quiescent galaxies at $z\sim 3$: 1) Stellar mass $M_\star > 10^{10.5} M_\odot$; 2) Photometric redshift ($2.5\leq z \leq 4.0$) ; 3) Spitzer MIPS 24$\mu m$ undetected (We cross-match our pVJL sample with the deep Spitzer MIPS 24$\mu m$ catalogue by Le Floc'h et al. 2009); 4) Specific star formation rate $sSFR=\frac{SFR}{M_\star}<10^{-9.5}$ yr$^{-1}$; 5) X-ray undetected by C-COMSOS (Civano et al. 2012); 6) Effective radius $r_e< 1"$; For the structural parameters of our pVJL galaxies, we use the results of a public catalogue including the structural parameters of the best-fitting \ser models of galaxies in CANDELS, given by van der Wel et al. (2012)\footnote{ftp://cdsarc.u$-$strasbg.fr/pub/cats/J/ApJS/203/24/}. Throughout this letter, we use the structural parameters measured using $\mh$ band images. 7) $\mh\leq 24.5$. Finally, we obtain 19 galaxies fulfilling all the conditions (See the filled points in Figure 1). We take them as our massive, quiescent galaxy candidates at $z\sim 3$ for further analysis.

\section{The structure of massive pVJL galaxies at z$\sim$3}

\begin{figure}
\plotone{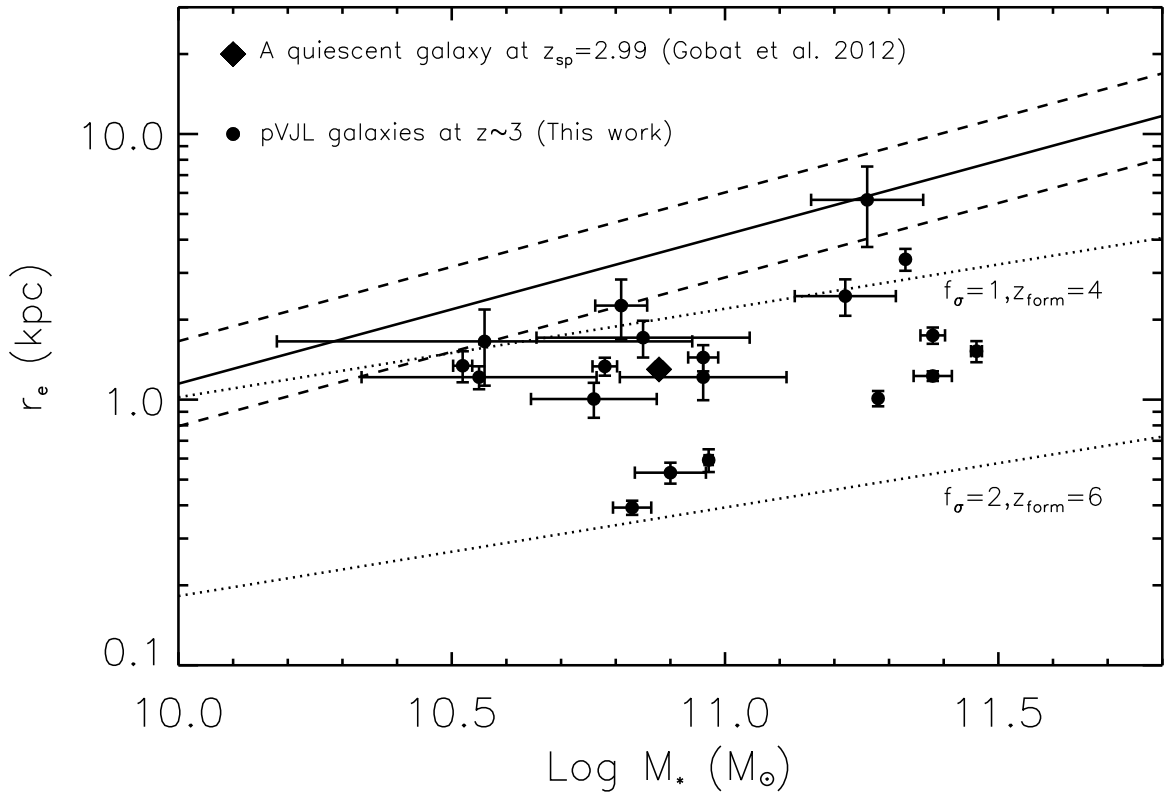}
\caption{The distribution of effective radius versus stellar mass for
  pVJL galaxies at $z>2.5$. Our pVJL sample is
  represented by black filled circles. A massive, quiescent galaxy with spectroscopic redshift
  $z_{spec}=2.99$ (Gobat et al. 2012) is represented by the filled diamond. The solid line shows the local
  effective radius and stellar mass relation of early-type
  galaxies (Shen et al. 2003). The dashed lines indicate the $\pm 1
  \sigma$ scatter of Shen et al. (2003) relation. The dotted lines illustrates the outcomes of model
  prediction (Fan et al. 2010) for reasonable values of the relevant parameters $f_\sigma$ and $z_{form}$ (see text).}
\end{figure}

In Figure 2, we plot the distribution of effective radius versus
stellar mass for our pVJL sample at $2.5\leq z \leq 4.0$.
We compare with the local relation  (Shen et al. 2003) and find that
these galaxies are on average $\sim 3-4$ times smaller than the local
counterparts with similar stellar masses. In Figure 2, we over-plot a massive and quiescent galaxy with spectroscopic
redshift $z_{spec}=2.99$, which is serendipitously discovered  by Gobat et al. (2012).
This spectroscopic-confirmed quiescent galaxy has very similar size as the median value of
our pVJL galaxies.
The compact size of these  $z\sim3$ galaxies can be modelled by assuming their formation according to the dissipative
collapse of baryons. The dotted lines  illustrates the outcomes of our model prediction  for reasonable values of the relevant parameters $f_\sigma$ and $z_{form}$ , where $z_{form}$ is the redshift when the collapse of baryons begins, and $f_\sigma$ is a factor relating the halo rotational velocity to the 3-D stellar velocity dispersion (see more details in Fan et al. 2010). For our pVJL galaxies at $z\sim3$ with luminosity-weighted ages $\geq 0.5$ Gyr, we expect their formation redshift should be $z_{form}\sim 4-6$.

\begin{figure}
\plotone{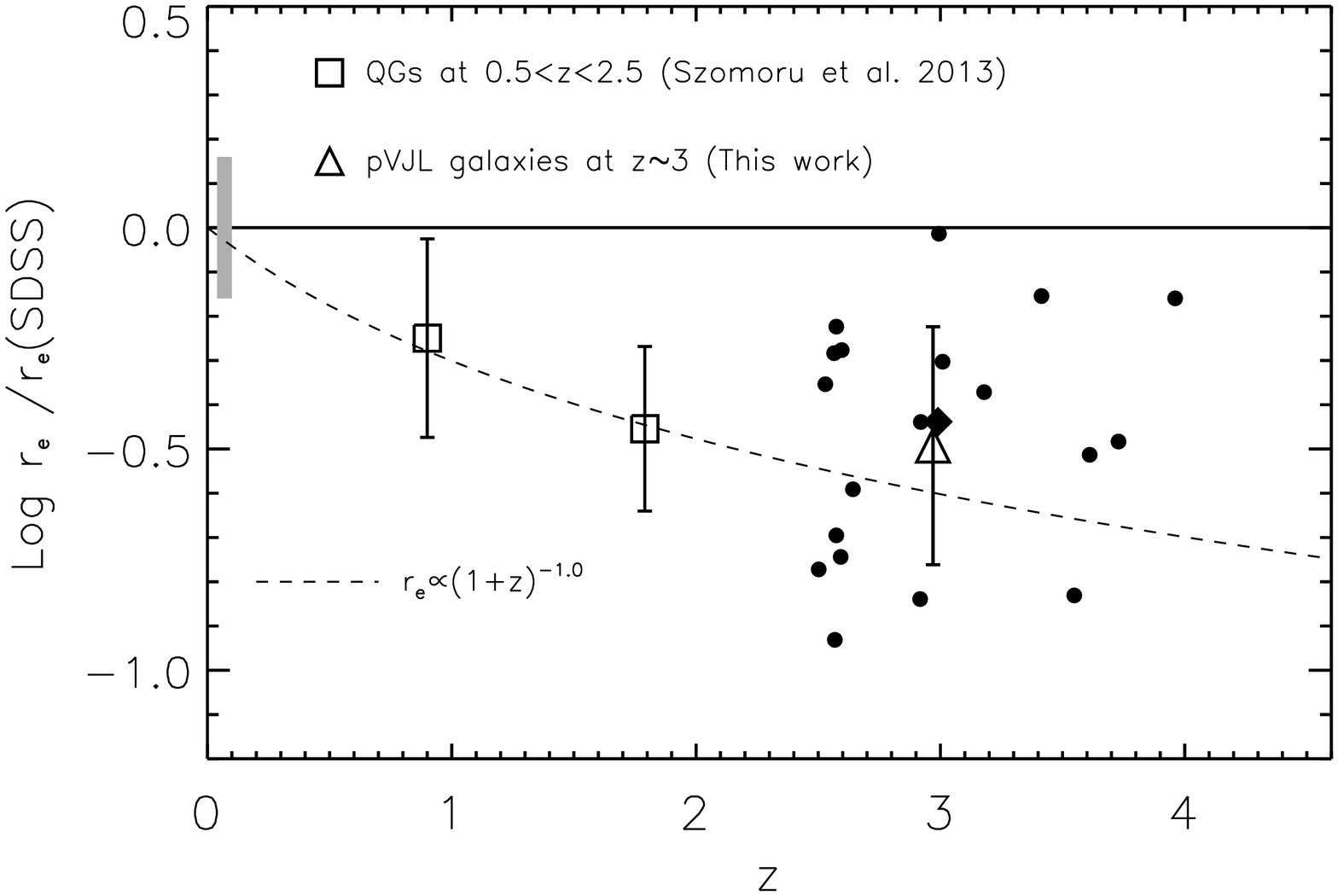}
\caption{Evolution of the effective radius with redshift since $z>3$. The shaded
  area reflects the distribution of local SDSS galaxies.
  The triangle symbol and error bar show the median size and $1 \sigma$ scatter of our pVJL sample at $z\sim 3$.
  The two squares show the median size of quiescent galaxies (QGs) at $0.5<z<2.5$ taken from Szomoru et al. (2013).
  The lines represents the size evolution with redshift described by $r_e\propto
   (1+z)^{\beta}$, with $\beta= -1.0$ for dashed line.
   Other symbols are the same as those in Figure 2. }
\end{figure}

In Figure 3, we explore the size evolution with redshift for our pVJL sample at $z\sim3$.
We determine the ratio between the measured effective radius of our pVJL galaxy ($r_e$) and the average
size with the same stellar mass expected by the local size-mass relation ($r_e(SDSS)$).
We adopt the local size-mass relation $r_e \propto M_\star^\alpha$, where $\alpha$ equals 0.56
by Shen et al. (2003). The relation $r_e\propto  (1+z)^\beta$ is used to describe the mass-normalized
size evolution with redshift. Combined with local SDSS data and CANDELS-GOODS-South data at $0.5<z<2.5$ (Szomoru et al. 2013),
the mass-normalized size evolution can be described well with $\beta\sim-1.0$ up to $z\sim4$.
The $1\sigma$ size scatter ($\sigma_{{\rm log}{r_e}}$) of our pVJLs sample is large ($\sigma_{{\rm log}{r_e}} \sim 0.27$).
These results are consistent with the previous findings based on the data at $z\le2.5$ (e.g. Newman et al. 2012; Szomoru et al. 2012; Fan et al. 2013).

The \ser indices can determine the distribution of light and can be suggestive of a bulge or disk component. The axis ratio distribution provides a better constraint on the shapes of galaxies. We examine \ser index of our pVJL galaxies taken from van der Wel et al. (2012). The \ser index of our pVJL galaxies at $z\sim3$ are relatively lower compared to the $z\sim2$ QGs. The median \ser index n is 1.5. And 13 out of 19 galaxies have $n<2.0$. The large fraction with low n indicates that the structure of our pVJL galaxies at $z\sim3$ may be mostly disk-dominated, same as the situation at $z\sim2$ (van der Wel et al. 2011). We also examine the axis ratios of our pVJL galaxies at $z\sim3$. The median axis ratio is $b/a\sim 0.65$. 9 out of 19 pVJL galaxies have $b/a< 0.6$, also indicating a disk-like structure.


\section{Stellar mass surface density profiles of massive pVJL galaxies at z$\sim$3}

In this section, we will estimate the stellar mass surface density profiles
of our pVJL galaxies at $z\sim3$ and compare it with those at lower redshift.
For simplicity, we neglect the radial gradients in the mass-to-light ratio and
use the best-fitting \ser profile representing the mass profile for each galaxy.
A \ser function is written as
\begin{equation}
\Sigma(r)=\Sigma_e exp(-b_n [(r/r_e)^{1/n}-1])
\end{equation}
Here $\Sigma(r)$ is the surface brightness at radius $r$, $r_e$ is the effective
radius, $\Sigma_e$ is the surface brightness at radius $r_e$ and $n$ is the \ser index.
The constant $b_n$ can be determined from the condition that the luminosity inside
$r_e$ is half the total luminosity, which has a
numerical solution (see Prugniel \& Simien 1997):
\begin{equation}
b_n = 2n - \frac{1}{3} + 0.009876/n
\end{equation}
Neglecting the radial gradients in the mass-to-light ratio, the stellar mass inside
radius r can be written as:
\begin{equation}
M(<r)=\gamma(2n,b_nx^{1/n}) \times M_\star
\end{equation}
where $x=r/r_e$, $M_*$ is stellar mass of the whole galaxy,
$\gamma(a,b)=(\int^{b}_{0}e^{-t}t^{a-1}dt)/(\int^{\infty}_{0}e^{-t}t^{a-1}dt)$
is the incomplete $\gamma$ function.
So basically, we can compute the stellar mass surface density profile
if having only effective radius $r_e$, \ser index n and stellar mass $M_*$.
For 19 pVJL galaxies at $z\sim3$, we compute the individual stellar mass surface
density profiles and the corresponding median profile at each radius. Our result is plotted in Figure 4 with solid lines
(thick line: median profile; thin lines: individual profiles).
We compare the median mass profile of our pVJL galaxies with those of QGs at $z\sim1$ and
$z\sim2$ (Szomoru et al. 2013) and local ETGs (Huang et al. 2013).
For the sake of comparison, we re-calculate the mass profiles of quiescent galaxies
in Szomoru et al. (2013) by neglecting radial gradients in the mass-to-light ratio.
In Figure 4, we can find that the median mass profiles at different redshifts
are very similar in the inner region (within $r<1-2$ kpc).
Anyway, we need notice that the median mass profiles are very uncertain
at $r<\sim0.7$ kpc, which is the approximate PSF HWHM at $z\sim 3$.
Despite the similarity, the median mass profiles at different redshifts already show some
difference at $1-2$ kpc. This difference is likely due to the large uncertainty of the median profiles
in the small samples at $z\ge1$. The difference is greater at larger radii ($r\geq 4$ kpc).
Figure 4 shows clearly the surface mass density increases with decreasing redshift at $r\geq4$ kpc.
The effective radii (marked with the colorful arrows) also increase with decreasing redshift.
One possible indication is most of stars in the inner region of local massive ETGs have been presented in place even since $z\sim3$.
This result is consistent with the inside-out formation scenario
in which the compact galaxies at high redshift make up the cores of local massive
ETGs and then build up the outskirts according to dissipationless minor mergers (e.g., Loeb \& Peebles 2003, Hopkins et al. 2009).
Patel et al. (2013) also studied the evolution in the surface mass density profiles by using CANDELS HST WFC3 imaging in the UKIDSS-UDS.
They confirmed that most of the mass at $r<2$ kpc was in place by $z\sim2$, and
that most of the new mass growth occurred at larger radii. However, they found the stellar mass appears to grow in the inner part of the galaxy
from $z\sim3$ to $z\sim2$. The difference of the evolution in the surface mass density profiles from $z\sim3$ to $z\sim2$ may be due to
the different sample selection methods. We compare the median mass profiles at different redshifts
with \emph{equal} stellar mass ($M_\star\sim 10^{11}M_\odot$), while Patel et al. (2013) selected the massive galaxies
at constant cumulative number density.  Their subsample at $z>2.5$ has a stellar mass of $10^{10.7}M_\odot$, which is two times less massive than ours.

\begin{figure}
\plotone{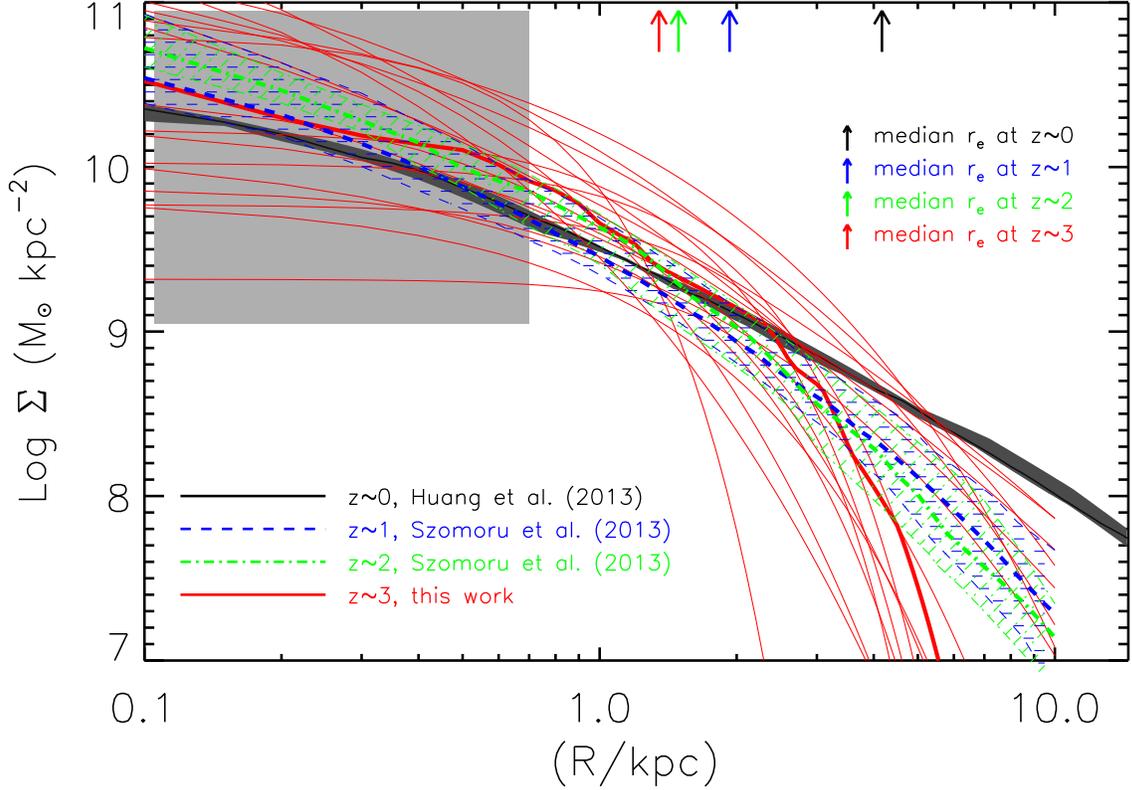}
\caption{The median stellar mass surface density profiles of massive, quiescent galaxies at
different redshift bins. The thick solid line shows the median profile of our pVJL sample at $z\sim 3$.
The individual profiles for the whole sample are plotted as thin solid lines.
The black shaded area represents the median profile and the associated $1 \sigma$ uncertainty range
of local massive early-type galaxies (Huang et al. 2013). We also plot the median profiles for massive, quiescent galaxies in CANDELS-GOODS-South field at $z\sim 1$ and $z\sim 2$, with the dashed and dot-dashed lines respectively (Szomoru et al. 2013). And the associated $1 \sigma$ uncertainty ranges are marked with the corresponding shaded areas. For the sake of comparison, we re-calculate the mass profiles of quiescent galaxies in Szomoru et al. 2013 by neglecting radial gradients in the mass-to-light ratio. The approximate PSF HWHM at $z\sim 3$ is indicated by the gray shaded region. The median effective radii ($r_e$) at different redshift bins are marked with the colorful arrows
(black: $z\sim0$; blue: $z\sim1$; green: $z\sim2$ and red: $z\sim3$ ).}
\end{figure}

\section{Summary and Discussion}

In this letter, we try to use a two-color (($J-L$) vs. ($V-J$)) selection criteria (so-called pVJL selection, Guo et al. 2012) to select massive, quiescent galaxy candidates at $2.5\leq z \leq 4.0$ in the CANDELS-COSMOS field.
The preliminary pVJL sample contains a lot of contaminations, which are mostly the highly obscured star forming galaxies  from both lower ($z\leq 2.0$) and higher ($z\geq 4.0)$ redshifts. And AGNs are another possible contamination, though they can only contribute a tiny fraction. Following the work in Guo et al. (2012), we use several conditions to clean our pVJL sample. Finally, 19 pVJL galaxies have been selected fulfilling all our massive, quiescent galaxy conditions at $z\sim3$ .
The caveat is there may be the remaining contamination from dusty SFGs in our pVJL sample. Though we try to clean sample by using available data, such as Spitzer 24 $\mu m$ detection, specific star formation rate based on the SED best-fitting result, it's still not adequate due to the poor limit of shallower 24$\mu m$ on star formation rate and the age-attenuation degeneracy of SED fitting. Further deeper submm/mm observation may help purify the quiescent sample.

By comparing our pVJL galaxies with local ETGs in the mass-size plane, we find that the sizes of our pVJL galaxies are on average 3-4 times smaller than those of the local ETGs with analogous stellar mass. The compact size of these  $z\sim3$ galaxies can be modelled by assuming their formation at $z_{form}\sim 4-6$ according to the dissipative collapse of baryons.

We extend our knowledge on the size evolution with redshift for massive, quiescent galaxies beyond $z\sim3$. The effective radius seems to gradually decrease with increasing redshift. Up to $z<4$, the mass-normalized size evolution can be described by $r_e\propto  (1+z)^{-1.0}$. The large scatter of size distribution $\sigma_{{\rm log}{r_e}}$ is also present in our pVJL sample at $z\sim3$. The remaining contamination from dusty SFGs may be also possibly corresponding to the large size scatter at $z\sim3$. Or repeated AGN feedback may help explain the increasing size scatter with redshift (e.g. Fan et al. 2013). By examining the \ser index and axis ratios , we find a significant fraction of our pVJL galaxies is disk-dominated. We need mention that the size comparison between massive quiescent galaxies with \emph{equal } stellar mass at $z\sim3$ and at $z\sim2$ may be not straightforward here. The progenitors of $z\sim2$ massive compact galaxies are compact star-forming galaxies at $z\sim3-4$ (Barro et al. 2013; Stefanon et al. 2013), which are excluded in this work according to the sSFR selection criteria. And our pVJL galaxies could be the possible progenitors of even more massive quiescent galaxies at $z\sim2$.

By comparing the median stellar mass surface density profile of our pVJL galaxies with those of QGs at $z\sim1$ and
$z\sim2$ and local ETGs.  An important result is most of stars in the inner region ($r< 1-2$ kpc) of local massive ETGs have been presented in place as early as $z\sim3$. The result favors the inside-out formation scenario in which the compact galaxies at high redshift make up the cores of local massive ETGs and then build up the outskirts according to dissipationless minor mergers.

\acknowledgments

We thank the referee for the careful reading and the valuable comments that helped improving our paper.
We also thank Dr. S. Huang, Dr. D. Szomoru and Dr. E. Le Floc'h for kindly providing us the electronic data from their works.
This work is supported by the National Natural Science Foundation of
China (NSFC, Nos. 11203023, 11225315) and Chinese Universities Scientific
Fund (WK2030220011,WK2030220004,WJ2030220007). L.F. thanks the partly
financial support from the China Postdoctoral Science Foundation (Grant
No.:2012M511411) and X.K. thanks the partly financial support from the
Doctoral Program of Higher Education (SRFDP, No. 20123402110037).

\textit{Facilities}: HST (WFC3, ACS)

\end{document}